\documentclass{article}
\usepackage{spconf,amsmath,graphicx}
\usepackage{cite}
\usepackage{latexsym}
\usepackage{amsfonts,amssymb}
\usepackage[hyperfootnotes=false]{hyperref}
\usepackage{algorithmic}
\usepackage{array}
\usepackage{stfloats}
\usepackage{url}
\usepackage{lipsum}
\usepackage{upgreek}
\usepackage{hhline}
\usepackage{enumerate}
\usepackage{mathtools}
\usepackage{booktabs}
\usepackage{lscape}

\usepackage{multirow}
\usepackage{colortbl}
\newcolumntype{C}[1]{>{\centering\arraybackslash}p{#1}}

\usepackage{booktabs,siunitx}
\usepackage{tabularx}
\usepackage[none]{hyphenat}%%%%

\newcommand{\befSecSpace}{\vspace{-0.5em}}
\newcommand{\afSecSpace}{\vspace{-0.5em}}
\newcommand{\befSubSecSpace}{\vspace{-1em}}
\newcommand{\afSubSecSpace}{\vspace{-0.5em}}

\usepackage{enumitem}% http://ctan.org/pkg/enumitem

\title{UX-Net: Filter-and-Process-based Improved U-Net for Real-time Time-domain Audio Separation}
\name{Kashyap Patel, Anton Kovalyov, and Issa Panahi}
\address{Electrical and Computer Engineering, University of Texas at Dallas, Richardson, TX, USA}
\begin{document}\sloppy
\ninept% For spacing
\maketitle
\begin{abstract}
This study presents UX-Net, a time-domain audio separation network (TasNet) based on a modified U-Net architecture. The proposed UX-Net works in real-time and handles either single or multi-microphone input. Inspired by the filter-and-process-based human auditory behavior, the proposed system introduces novel mixer and separation modules, which result in cost and memory efficient modeling of speech sources. The mixer module combines encoded input in a latent feature space and outputs a desired number of output streams. Then, in the separation module, a modified U-Net (UX) block is applied. The UX block first filters the encoded input at various resolutions followed by aggregating the filtered information and applying recurrent processing to estimate masks of separated sources. The letter `X' in UX-Net is a name placeholder for the type of recurrent layer employed in the UX block. Empirical findings on the WSJ0-2mix benchmark dataset show that one of the UX-Net configurations outperforms the state-of-the-art Conv-TasNet system by 0.85 dB SI-SNR while using only 16\% of the model parameters, 58\% fewer computations, and maintaining low latency.

\end{abstract}
\befSecSpace
\section{Introduction} \label{sec:intro}
\afSecSpace
% Hands-free voice-assisted technologies have seen tremendous growth in recent years. Speech-to-text services, smart home assistants, and automatic meeting diarization are just a few examples. One well-known issue with these technologies is that they are susceptible to errors when faced with a multi-talker scenario. Possible solutions include extracting the speech of the targeted speaker \cite{speaker_ext} (Speaker Extraction), separating all overlapping speech from the mixture \cite{bss} (Speech Separation), and attending to the information separately. Speech separation is a speaker-independent and more generic technique that piqued academic curiosity. Recent developments in deep learning models have significantly improved the performance of state-of-the-art (SOTA) speech separation models on various benchmark datasets \cite{supervised_overview, comprehensive_time_freq}. The above mentioned speech applications are often required to be real-time and energy-efficient. On the other hand, deep learning-based (DL) systems are computationally expensive and memory-intensive. Considering these objectives, this paper provides a deep learning architecture for real-time (causal) speech separation with single or multi-channel input that is both computationally and memory-efficient.\par
Hands-free voice-assisted technologies have seen tremendous growth in recent years. Speech-to-text services, smart home assistants, and automatic meeting diarization are just a few examples. However, susceptibility to errors in multi-talker scenarios is a well-known limitation of these technologies. Solutions include: extracting the speech of the targeted speaker \cite{speaker_ext} (Speaker Extraction); and separating all overlapping speech from the mixture \cite{bss} (Speech Separation) followed by attending to the information separately. Speech separation is a speaker-independent and more generic technique that piqued academic curiosity. Recent developments in deep learning models have significantly improved the performance of state-of-the-art (SOTA) speech separation models \cite{supervised_overview, comprehensive_time_freq} on various benchmark datasets. Deep learning-based (DL) solutions are known to be computationally and memory demanding. However, speech processing applications are often constrained to run in real-time and be energy-efficient. Therefore, this paper offers a deep learning architecture for real-time (causal) speech separation, for either single or multi-channel input, that is both computationally and memory efficient.\par
% ; therefore, one must be cautious while designing a solution. 

%Neural networks are better at predicting probabilities or masks than generating raw signals. Following that approach, initial work in frequency-based DL includes estimating the magnitude spectrum of each speaker by masking the mixture magnitude spectrum using ratio masking targets. Mixture phase information was used to reconstruct the signals back in the time domain. Later multiple approaches were developed to target the phase of information retrieval.  

The time-domain audio separation network (TasNet) is a significant class among the best-performing DL systems. TasNet follows an encoder-decoder-based structure which transforms the time domain signal into a latent space, analogous to the short-time Fourier transform (STFT) domain, where it estimates masks of the sources followed by reconstruction of the separated signals into the time domain. In TasNet, permutation invariant training \cite{pit} (PIT) is employed to solve the permutation problem. The initially proposed non-causal bi-directional long-short term memory \cite{tasnet} (Bi-LSTM) based TasNet was shown to outperform STFT-based DL approaches \cite{comprehensive_time_freq}. Separation was then improved with the dilation-based temporal convolution network (TCN) in Conv-TasNet \cite{convtasnet}. The results of Conv-TasNet suggested that long-term sequential context awareness is needed to process auditory information effectively. As a result, dual-path data segmentation-based neural networks, such as the Dual-path Recurrent Neural Network \cite{dprnn} (DPRNN), the Dual-path Transformer Neural Network \cite{DPTNN} (DPTNN), and the Globally Attentive Locally Recurrent \cite{GALR} (GALR) network, were proposed to further improve separation by processing both local and global contexts. Inspired by the success in image segmentation, U-Net \cite{UNet} based architectures, such as Wave-U-Net \cite{waveUNet} and Sudo Rm-Rf \cite{SudoRmRf}, also became popular among TasNet-like systems. In a U-Net, a signal is repeatedly downsampled and upsampled with skip connections at different resolutions to provide extended context aggregation.\par

% It was then suggested that long-term sequential context awareness is needed to process auditory information effectively and performance was further improved with the dilation-based temporal convolution network (TCN) proposed in Conv-TasNet \cite{convtasnet}. Further improved separation was achieved by processing both local and global contexts in the Dual-path Recurrent Neural Network \cite{dprnn} (DPRNN), the Dual-path Transformer Neural Network \cite{DPTNN} (DPTNN), and the Globally Attentive Locally Recurrent \cite{GALR} (GALR) model.

% Efficiency and separation performance of TasNet was then improved with with the dilation-based temporal convolution network (TCN) in Conv-TasNet \cite{convtasnet}. It suggested long-term sequential context awareness is required to process auditory information effectively. Consequently, dual-path data segmentation-based neural networks such as Dual-path Recurrent Neural Network \cite{dprnn} (DPRNN), Dual-path Transformer Neural Network \cite{DPTNN} (DPTNN), and Globally Attentive Locally Recurrent Network \cite{GALR} (GALR) were proposed to process global and local contexts. These models achieved improved separation results. Inspired by the success in image segmentation, U-Net \cite{UNet} based architectures became popular in speech separation. Wave-U-Net \cite{waveUNet} and Sudo-RmRf \cite{SudoRmRf} are U-Net architecture-based speech separation models. In U-Net, the time domain signal is repeatedly downsampled and upsampled with skip connections at each level to provide extended context aggregation over multiple signal resolutions.\par

In real-time or causal speech separation, speech is separated using only current and past data. Few causal speech separation architectures have been proposed in the literature. Tweaking Conv-TasNet into a causal configuration resulted in significant performance deterioration when compared to its non-causal counterpart. The dilated temporal convolutions in Conv-TasNet keep a substantial amount of data history at each convolutional block, which, along with its high number of skip connections, makes the model highly memory inefficient. Due to dual-path processing, the local and global context-aware DPRNN, DPTNN, and GALR models are unsuitable for causal processing. Similarly, the U-Net-based architectures Wave-U-Net and Sudo Rm-Rf perform resampling in the time axis, making them inherently non-causal. Inspired by research in computational auditory scene analysis (CASA), the causal Deep CASA \cite{DeepCASA} presented a U-Net-based clustering algorithm for causal speech separation in the frequency domain. Although Deep CASA attains high separation performance, it incurs excessive latency when compared to TasNet-like systems.\par

% Moreover, causal operation limits resampling along the time axis in U-Net-based designs, reducing the performance of existing U-Net-based structures \cite{waveUNet, SudoRmRf}. Deep-CASA \cite{DeepCASA}, a frequency domain approach, presented a U-Net-based clustering algorithm that produced satisfactory causal separation results but at the expense of considerable processing cost and latency.\par

Motivated by the above observations, we revise the U-Net architecture and TasNet structure for low-cost, low-latency causal speech separation. As a result, we offer UX-Net. The human brain processes mixed sound at several resolutions and contexts, first masking out undesired noises, i.e., \textit{filtering}, then aggregating the information, and finally \textit{processing} the individual sounds in parallel \cite{brain-1, brain-2}. Inspired by this filter-and-process technique, our system introduces novel cost and memory efficient mixer and separation modules. In the mixer module, the encoded, either single or multi-channel input is mixed and mapped into a desired number of output streams. In the separation module, the mixer output is processed by a modified U-Net (UX) block. The initial half of the UX block filters the encoded input at various resolutions using convolutional neural network (CNN) units. The second half of the UX block employs a set of CNN and recurrent neural network (RNN) units to aggregate and process the filtered input at the different resolutions. In contrast to a typical U-Net, resampling is performed solely across the feature dimension to ensure causality. CNNs have a local receptive field and are excellent at filtering, while RNNs with gated capabilities, such as LSTM or Gated Recurrent Units (GRUs), provide an adaptive receptive field and longer context window without explicitly retaining long data history. Thus, the combination of both is used in UX-Net to improve upon the classical design of a U-Net. The letter ‘X’ in UX-Net is a name placeholder for the type of RNN employed.

UX-Net is benchmarked against SOTA causal speech separation methods using the WSJ0-2mix dataset. Results show that the proposed system is capable of achieving very high separation performance while incurring nearly negligible latency, and comparatively low memory and computational complexities. Furthermore, using a simulated dataset derived from LibriSpeech, we investigate the performance of UX-Net when faced with either single or multi-channel reverberant input. An ablation study is also conducted.\par

\befSecSpace
\section{Problem Formulation}
\afSecSpace

Let us consider a microphone array of $M$ microphones. Let $\mathbf{x}_m$ be a vector representing a finite acoustic time signal captured by the $m$-{th} microphone. $\mathbf{x}_m$ comprises a convolutive mixture of $C$ overlapping speech sources given by 
% \begin{equation}\label{eq:formulation}
%     \mathbf{x}_m = \sum_{i=1}^{C} \mathbf{s}_i \ast (\mathbf{h}_{mi}^{early} + \mathbf{h}_{mi}^{reverb}) \;, \quad m \in \{1, \ldots, M\} \; ,
% \end{equation}
\begin{equation}\label{eq:formulation}
    \mathbf{x}_m = \sum_{i=1}^{C} \mathbf{s}_i \ast \mathbf{h}_{mi} \;, \quad m \in \{1, \ldots, M\} \; ,
\end{equation}
where $\mathbf{s}_i$ is the utterance of the $i$-th speech source and $\mathbf{h}_{mi}$ is the impulse response of the $i$-th source with reference to the $m$-th microphone. For simplicity, background and internal microphone noises are neglected. The impulse response $\mathbf{h}_{mi}$ can be split into its early ($\mathbf{h}_{mi}^{early}$) and late ($\mathbf{h}_{mi}^{reverb}$) reflection components as given by
\begin{equation}
    \mathbf{h}_{mi} = \mathbf{h}_{mi}^{early} + \mathbf{h}_{mi}^{reverb} \;.
\end{equation} 
The late reflection component introduces reverberation, which may be undesired depending on application.\par
% Speech separation is a challenging problem in itself. Hence
Let the first microphone be the reference microphone. Speech separation is defined here as extracting all individual speech sources as captured by the reference microphone. Thus, the target is given by
% Pure speech separation with respect to reference microphone (${m_0}$) is defined as estimating all the overlapped speech sources from the available mixture recordings. It is defined as,
\begin{equation}\label{eq:T1}
    \mathbf{s}_{1 i} = \mathbf{s}_{i} \ast \mathbf{h}_{1 i}\;, \quad i \in \{1, \ldots, C\} \; .
\end{equation}
The target of speech separation with dereverberation, on the other hand, is defined as
\begin{equation}\label{eq:T2}
    \mathbf{s}_{1i} = \mathbf{s}_{i} \ast \mathbf{h}_{1i}^{early}\;, \quad i \in \{1, \ldots, C\} \; .
\end{equation}\par

A common approach in signal processing is to divide the signals into overlapping time frames and process them. Therefore, the different-channel utterances $\mathbf{x}_m$ are split into $K$ sequential overlapping frames of length $L$ samples and stacked across the channel dimension, resulting in the 3-dimensional (3D) tensor $\mathbf{X} \in \mathbb{R}^{M \times K \times L}$. Given $\mathbf{X}$, the task is then to estimate the tensor $\hat{\mathbf{S}} \in \mathbb{R}^{C \times K \times L}$ representing the corresponding extracted frames of the $C$ speech sources. 

% Let $\mathbf{x} \in \mathbb{R}^{M \times K \times L}$ be a tensor grouping the overlapped frames of $\mathbf{x}_m$ for all $m$.
% A common approach in signal processing is to divide the signals into overlapping time frames and process them. Suppose, $\mathbf{x}_m$ is of observation interval $T$ samples and is divided into frames of length $L$ samples and hop size of $H$ samples, we get total $K = \lfloor \frac{T-1}{H} \rfloor + 1$ frames and $\mathbf{x}_m \in \mathbb{R}^{K \times L}$. Concatenating all the input channels, we get input $\mathbf{x} \in \mathbb{R}^{M \times K \times L}$. Similarly, assuming there exist $C$ sources in the mixture, concatenating all the separated speech signals across a single dimension, we get output $\hat{\mathbf{s}} \in \mathbb{R}^{C \times K \times L}$. 

\begin{figure}[t!]
\begin{minipage}[b]{1.0\linewidth}
  \centering
  \centerline{\includegraphics[width=8.5cm]{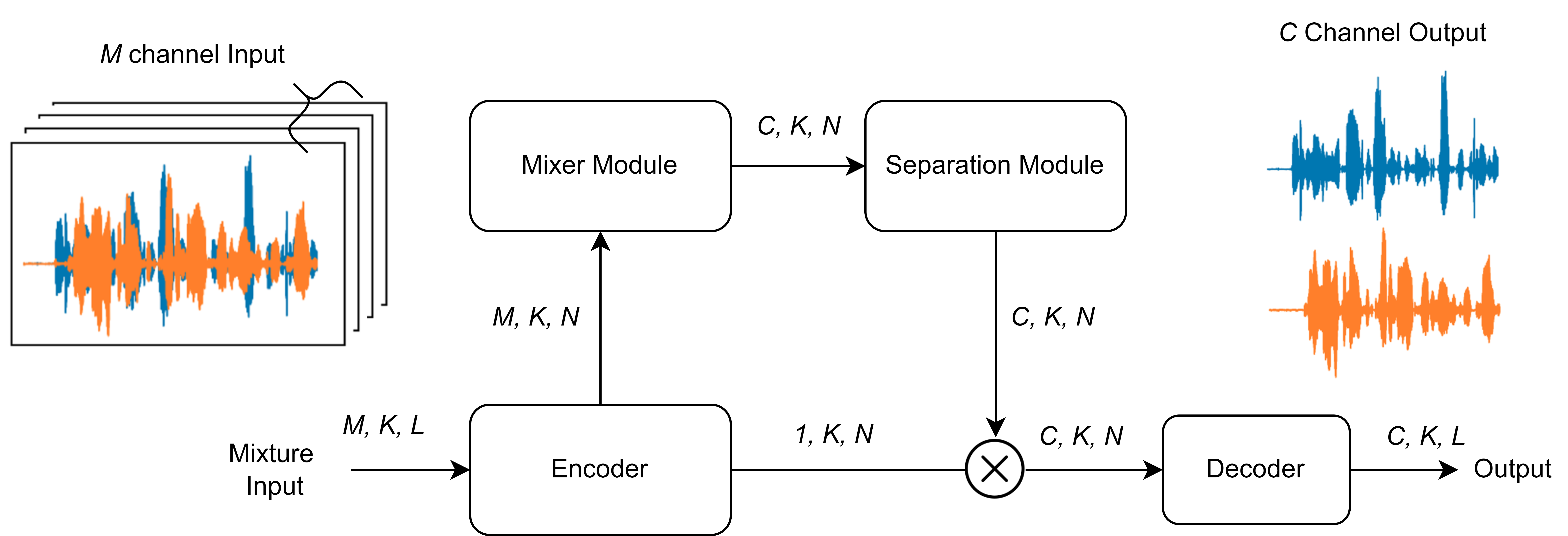}}
\end{minipage}
\caption{Block diagram of the proposed multi-channel speech separation pipeline. Example with $M=4$ and $C=2$.}
\label{fig:pipeline}
\vspace{-0.5cm}
\end{figure}
% The C channel output waveform diagram on the right is the proposed model's output example for a mixture input shown on the left.
\befSecSpace
\section{Separation System}
\afSecSpace
\par 
As shown in Fig. \ref{fig:pipeline}, the proposed separation system consists of encoder, mixer, separation, and decoder modules.\par 
% The introduced mixer module can be configured to work with various input channels and output sources. 
%Separation system receives raw time-domain frame-based multi-channel, $\mathbf{x}$ as input. 

\befSubSecSpace
\subsection{Encoder}
\afSubSecSpace
The encoder begins by cumulatively normalizing the raw time-domain input $\mathbf{X}$, which is then sent through a feed-forward layer $\mathcal{F}_e$ having weights $\mathbf{W}_e \in \mathbb{R}^{L \times N}$ and no bias. Thus, analogous to the STFT, $\mathcal{F}_e$ transforms each time-domain frame into its representation in a latent-space using the $N$ basis signals in $\mathbf{W}_e$. The complete encoder operation is given by 
\begin{equation}
    \mathbf{E} = \text{ReLU}(\mathcal{F}_e(\text{cLN}(\mathbf{X}))) \;,
    \label{eq:encoder}
\end{equation}
where $\text{ReLU}(\cdot)$ is the rectified linear unit and $\text{cLN}(\cdot)$ denotes cumulative normalization \cite{convtasnet}. The tensor $\mathbf{E} \in \mathbb{R}^{M \times K \times N}$ represents the weights of the normalized input mixture in a latent space. ReLU is used here to ensure non-negative weights.\par

\befSubSecSpace
\subsection{Mixer}
\afSubSecSpace
\par
% In the mixer module, the encoded $M$-channel input in $\mathbf{E}$ is mapped into
In the mixer module, the $M$-channel encoded input $\mathbf{E}$ is mapped onto a $C$-channel tensor by convolution across the time and feature axes followed by combination across the channel axis. As such, this module has two purposes: (1) learning spatial features for improved separation performance; (2) ensuring scalable model size for varying number of input channels. The mixer operation is given by
\begin{equation} \label{eq:mixer}
\begin{aligned}
\mathbf{E}_M &= \text{PReLU}(\text{cLN}(\text{Conv2D}_{M, M, (3 \times 3)}(\mathbf{E}))) \\
\mathbf{E}_C &= \text{PReLU}(\text{cLN}(\text{Conv2D}_{M, C, (3 \times 3)}(\mathbf{E}_M))) \;,
\end{aligned}
\end{equation}
where $\text{Conv2D}_{M_1, M_2, (3 \times 3)}(\cdot)$ is a 2D $3\times 3$ convolutional layer with $M_1$ and $M_2$ being the respective number of input and output channels, $\text{PReLU}(\cdot)$ denotes parametric ReLU \cite{PReLU}. $\mathbf{E}_M \in \mathbb{R}^{M \times K \times N}$ and $\mathbf{E}_C \in \mathbb{R}^{C \times K \times N}$ are respective intermediate and final outputs of the mixer. In all convolutional layers of UX-Net, the input is zero-padded from the left to ensure matching output size while preserving causality.\par
% Latent representation, $\mathbf{E}_C \in \mathbb{R}^{C \times K \times N}$ is a $C$ channel input to a separator module. $\text{Conv2D}_{M_1, M_2, (3 \times 3)}(\cdot)$ operator represents 2-dimensional $3\times 3$ convolutional layer with $M_1$ and $M_2$ as number of input and output channels respectively. All convolutional layers are zero-padded from left to maintain causality. $\text{PReLU}(\cdot)$ is a Parametric ReLU activation function \cite{PReLU}, used to introduce non-linearity. 

%Here the prefix $\text{DW}$ represents depth-wise convolution.

\begin{figure*}[t]
\begin{minipage}[b]{1.0\linewidth}
  \centering
  \centerline{\includegraphics[width=0.73\textwidth]{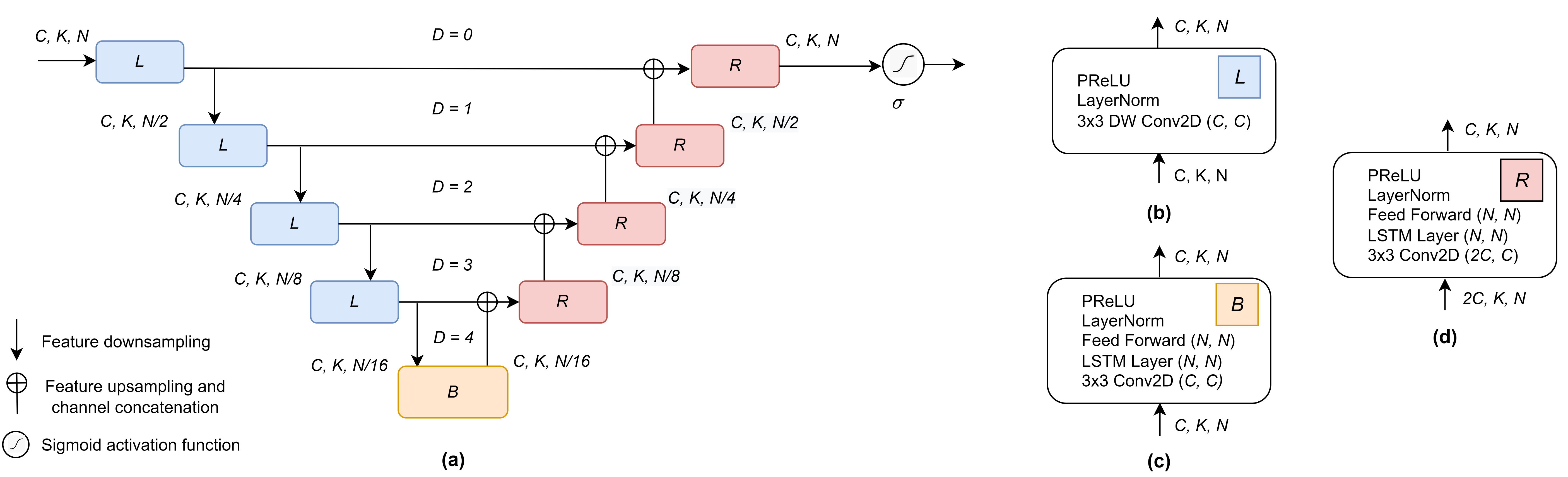}}
\end{minipage}
% \caption{(a) Separation module with one UL block (`X' := `L' due to use of LSTM layers) with depth $D = 4$ followed by Sigmoid non-linearity. (b) $L$ unit. (c) $B$ unit. (d) $R$ unit. }
\caption{Schematics of (a) a separation module with one UL block (`X' := `L' due to use of LSTM layers) with depth $D = 4$ followed by Sigmoid non-linearity, and corresponding (b) \textit{L}, (c) \textit{B}, and (d) \textit{R} units.}
\label{fig:UL-Net}
\vspace{-0.3cm}
\end{figure*}
%TODO: Mention dimension of RNN layers, in figures or in text. Figure is missing the dimensions to LSTM and FF layers. 

\befSubSecSpace
\subsection{Separation}
\afSubSecSpace
Given $\mathbf{E}_C$ in (\ref{eq:mixer}) as input, the separation module estimates the masks of each source with respect to the first channel of $\mathbf{E}$ in (\ref{eq:encoder}). Separation (in a latent space) is then achieved by
\begin{equation}
    \mathbf{E}_{S}^{(i)} = \mathbf{E}_{mask}^{(i)}\odot \mathbf{E}^{(1)}, \quad i \in \{1, \ldots, C\} \; ,
\end{equation}
where $\mathbf{E}^{(1)} \in \mathbb{R}^{1 \times K \times N}$ is the first channel of $\mathbf{E}$, the $\odot$ denotes the Hadamard product, $\mathbf{E}_{mask}^{(i)} \in \mathbb{R}^{1 \times K \times N}$ is the $i$-th channel of the estimated mask tensor $\mathbf{E}_{mask} \in \mathbb{R}^{C \times K \times N}$, and correspondingly $\mathbf{E}_{S}^{(i)} \in \mathbb{R}^{1 \times K \times N}$ is the $i$-th channel of the tensor $\mathbf{E}_{S} \in \mathbb{R}^{C \times K \times N}$ representing the separated sources in the latent space.\par
The proposed separation module consists of a UX block followed by a Sigmoid non-linearity $\sigma$. As such, mask estimation is given by
\begin{equation}
    \mathbf{E}_{mask} = \sigma (\text{UX}(\mathbf{E}_C)) \;.
\end{equation}
As depicted in Fig. \ref{fig:UL-Net}, a UX block is made up, sequentially, of $D$ left ($L$), one bottom ($B$), and $D$ right ($R$) units. $D$ is a parameter denoting the depth of the UX block. The $L$ unit filters the input using depth-wise (DW) convolution \cite{DW_convolution}, which avoids channel interaction (CI). Then, the output of the $L$ unit is downsampled across the feature dimension by a factor of two and fed as input to the next $L$ or $B$ unit in the sequence. The $B$ unit is applied at the lowest feature resolution. Both $B$ and $R$ units aggregate and process multi-channel input using a sequence of convolutional, recurrent, and feed-forward layers. The recurrent and feed-forward layers are applied in parallel across each channel. Parallel processing across the different channels limits computational complexity and enables shared weights, resulting in fewer parameters to train. The output of $B$ and $R$ units is upsampled, concatenated with matching size output of an $L$ unit and fed to the next $R$ unit in the sequence, up until the point the original feature resolution $N$ is recovered. Concatenation with matching-size output of an $L$ unit helps preserve information at a given resolution. Due to different-sized inputs, the difference between $B$ and $R$ units lies in that the convolutional layer of the former maps $C$-channel input onto $C$-channel output, whereas that of the latter maps $2C$-channel input onto $C$-channel output. Depending upon the type of recurrent processing applied in the $B$ and $R$ units, UX-Net is renamed to either UL-Net, if LSTMs are used, or UG-Net, if GRU layers are used.\par

The architecture of the proposed UX block is largely inspired by the filter-and-process-based human auditory behavior. Compared to the classical U-Net design, a UX block introduces three key differences: (1) resampling is performed across the feature axis instead of the time axis, thus preserving causality; (2) recurrent layers are introduced for global context awareness; (3) channel dimensionality is fixed instead of being doubled as resolution decreases, limiting the overhead of depth on the model complexity.\par

\befSubSecSpace
\subsection{Decoder}
\afSubSecSpace
The decoder transforms the extracted sources in the latent space to the corresponding time domain signals. First, $\mathbf{E}_S$ is mapped onto $\hat{\mathbf{S}}$ by a feed forward layer $\mathcal{F}_d$ with weights $\mathbf{W}_d \in \mathbb{R}^{N \times L}$ as given by
\begin{equation}\label{eq:output}
    \hat{\mathbf{S}} = \mathcal{F}_d(\mathbf{E}_{S}) \;.
\end{equation}
Then, the overlap-add method is applied on the $C$ channels of $\hat{\mathbf{S}}$ to extract the final separated waveforms.\par

\befSecSpace
\section{Experiments}
\afSecSpace
\subsection{Datasets}
% Separation performance of UX-Net is evaluated on two popular datasets.\par
We conduct speech separation experiments using two popular datasets: Wall Street Journal (WSJ) \cite{wsj} and LibriSpeech \cite{librispeech}.\par

Dataset 1: We consider the WSJ0-2mix \cite{deep-clustering, utterance-PIT, tasnet} two-speaker speech mixture dataset derived from WSJ. WSJ0-2mix is used for benchmarking UX-Net against different SOTA causal speech separation models. This dataset, however, has two limitations: (1) mixtures consist of close-talk speech lacking reverberant components; (2) only mixtures with 100\% overlap are considered, which are rare in practice.\par

Dataset 2: Aiming to overcome the limitations of WSJ0-2mix, we use clean speech utterances from LibriSpeech and generate a dataset simulating two overlapping speech sources captured by a microphone array in a reverberant room. This dataset contains 30 h training, 5 h validation, and 5 h test sets consisting of 4-second-long utterances. The signals are sampled at 8 kHz. For each utterance, the dimensions of the room are randomly sampled between 5 and 10 meters in length and width, and 2 to 5 meters in height. The reverberation time ranges randomly between 0.1 and 0.5 seconds. The speech overlap ratio and the signal-to-noise ratio (SNR) vary randomly between 5\% and 95\% and 0 and 5 dB, respectively. Speech sources are distributed randomly around the room with the constraint of being at least 50 cm away from the walls. The microphone array consists of a 5-element circular array with a radius of 5 cm. The microphone array is placed in the middle of the room and the image method \cite{habets-RIR} is applied to generate the corresponding room impulse responses (RIRs). The training target utterances include up to 50 ms of reverberation following the direct path \cite{reverb}, emphasizing separation and dereverberation as per (\ref{eq:T2}).\par

\befSubSecSpace
\subsection{Training and Evaluation}
\afSubSecSpace

Training loss is given by the negative of scale-invariant SNR (SI-SNR) averaged across the separated sources in a mixture. SI-SNR \cite{backed-sdr} measures the scale-invariant similarity between a target signal $\mathbf{s}$ and an estimated signal $\hat{\mathbf{s}}$. It is given by
\begin{equation}
    \text{SI-SNR} := 10\log_{10}\left(\frac{\| \alpha \mathbf{s} \|^{2}}{\| \hat{\mathbf{s}} - \alpha \mathbf{s} \|^{2}}\right) \; ,
\end{equation} 
where $\alpha =  \frac{ \hat{\mathbf{s}}^{T}\mathbf{s}}{\| \mathbf{s} \|^{2}}$ is the scalar projection of $\hat{\mathbf{s}}$ onto $\mathbf{s}$. PIT training with Hungarian algorithm \cite{hungarian} is employed to solve the permutation problem. The network is trained for 100 epochs with Adam \cite{adam} optimizer and a batch size of 4. The initial learning rate is set to $10^{-3}$ and later multiplied by 0.98 every two epochs. Gradients are clipped to ([-5, 5]) during the backward pass to avoid the exploding gradient problem. The models with best validation loss are saved and evaluation results are reported on the test sets. Consistent with other TasNet systems, we found that best performance is achieved using a small frame size. Thus, a frame size of 2 ms with 1 ms (50\%) overlap is used, resulting in a total algorithmic latency of only 3 ms when allowing 1 ms of processing time per frame.\par

The performance metrics used are: improvement in SI-SNR (SI-SNRi), Perceptual Evaluation of Speech Quality \cite{PESQ} (PESQ), and Short-Time Objective Intelligibility \cite{STOI} (STOI).\par

\begin{table}[h]
\vspace{-0.2cm}
 \caption{Comparison with causal SOTA on WSJ0-2mix dataset.}
%  Corresponding optimal values are highlighted in bold.
 \small
    \centering
    \setlength\tabcolsep{1.5pt} % default value: 6pt  
    \begin{tabular}{p{10.3em} p{5em} p{4em} p{7em}}  
            \toprule
                           Model      & Parameters (M) & FFPF \quad (M) & SI-SNRi(dB) /PESQ /STOI \\
             \midrule
      %                     &  Mixture                        & -                & -                & 0.00/ 2.02/ 0.56 \\
LSTM-TasNet  \cite{tasnet}                      & 32               & -                & 10.80/ -/ -    \\
Conv-TasNet \cite{convtasnet}                   & 5.05             & 5.23             & 10.60/ -/ -    \\
$\text{Deep-CASA}^{\diamond}$ \cite{DeepCASA}   & 12.8             & -                & 15.20/ 3.25/ 0.90    \\
\cline{1-4}
$\text{UG-Net}$ \phantom{,x} ($N=128$)          & 0.16             & 0.47             & 9.73/ 2.71/ 0.87   \\  
$\text{UG-Net}$ \;\;\; ($N=256$)                & 0.63             & 1.82             & 11.13/ 2.82/ 0.88   \\
$\text{UL-Net}$ \phantom{1x} ($N=128$)          & 0.20             & 0.56             & 10.18/ 2.78/ 0.87   \\
$\text{UL-Net}$ \phantom{1x} ($N=256$)          & 0.80             & 2.17             & 11.45/ 2.92/ 0.89   \\
$\text{UL-Net 2x}$ ($N=256$)                    & 1.59             & 4.32             & 12.41/ 3.01/ 0.90   \\
$\text{UL-Net 4x}$ ($N=256$)                    & 3.17             & 8.63             & 13.60/ 3.12/ 0.90   \\
             \bottomrule 
              \multicolumn{4}{p{0.475\textwidth}}{${}^{\diamond}$ Frequency domain method with a frame size of 32 ms. The rest of the methods have a frame size of 2 ms.}  \\
        \end{tabular}
        \label{table:wsj}
\vspace{-0.2cm}
\end{table}

\befSubSecSpace
\subsection{Results}
\afSubSecSpace

In the first experiment, we employ the WSJ0-2mix benchmark dataset to perform single-channel speech separation as defined in (\ref{eq:T1}). Table \ref{table:wsj} compares the performance of UX-Net to the SOTA causal time-domain LSTM-TasNet and Conv-TasNet, as well as the causal frequency-domain Deep CASA. Different configurations of UX-Net are considered by varying the recurrent layer used (either LSTM or GRU) and the latent space dimensionality $N$. We also evaluate the effect of deepening the separation module of UX-Net using multiple successive UX blocks with skip connections. The $n$ in UX-Net $n$x denotes the number of repeated UX blocks. The depth $D$ is set to 5 in all UX-Net configurations. For the SOTA methods, only the best reported results are listed, and the unreported fields are left blank. The field FFPF stands for \textit{forward floating-point operations per frame} and is used to measure the per-frame computational burden during inference. The results show that UX-Net can outperform both LSTM-TasNet and Conv-TasNet while incurring significantly lower computational and memory cost. In fact, the LSTM-based configuration with $N = 256$ outperforms Conv-TasNet by 0.85 dB SI-SNRi while needing only 16\% of the model parameters and 58\% fewer computations. Deep-CASA attains the best performance, but not without incurring somewhat excessive system latency common to frequency-domain models. Separation performance of UX-Net is shown to improve as the number of UX blocks increases.\par

In the second experiment, we employ the generated multi-channel LibriSpeech dataset to perform speech separation with dereverberation as defined in (\ref{eq:T2}). Table \ref{table:multichannel} compares the performance of UX-Net to Conv-TasNet. By default, the number of input channels is one unless otherwise specified. In all UX-Net configurations, we let $D=5$ and $N=256$. For Conv-TasNet, we used the author's best-performing implementation and trained it with the generated dataset under the same conditions as UX-Net. It is confirmed once more that the proposed system outperforms Conv-TasNet. Moreover, we notice that the performance of UX-Net improves for an increased number of input channels without a significant effect on computational and memory complexities.\par

In the third experiment, we conduct an ablation study of UX-Net to verify the effectiveness of increased depth, no CI in $L$ units, and use of cLN. Table \ref{table:abalation} reports the results. RTF gives the \textit{real-time factor} of an AMD Ryzen 7 3800X CPU when processing one second of input data using a specified model configuration. We see that increasing the depth improves performance without significant effect on RTF. In contrast to a typical U-Net, little effect on RTF is attributed to the fact that as depth increases, the number of additional parameters needed in UX-Net decreases exponentially. We also notice that when CI is allowed by substituting the depth-wise convolution in an $L$ unit with regular convolution, SI-SNRi drops by 0.11 dB. Furthermore, replacing cLN with framewise normalization lowers SI-SNRi by 0.54 dB. Finally, low RTF values confirm the real-time feasibility of UX-Net.\par

\begin{table}[t]
% \vspace{-0.2cm}
\caption{Analysis on reverbarant multi-channel LibriSpeech dataset.}
\small
\centering
\setlength\tabcolsep{1.5pt} % default value: 6pt
    \begin{tabular}{p{10.3em} p{5.3em} p{4em} p{6em}}
        \toprule
         Model & Parameters (M) & FFPF \quad (M) &{SI-SNRi(dB)} /{PESQ}/{STOI}   \\
         \midrule
         $\text{Conv-TasNet}$                 & 5.05  &5.23 & 6.57/2.15/0.75                 \\ 
         \midrule
         UL-Net                               & 0.80  &2.17 & 7.35/2.27/0.79                \\ 
         UG-Net                               & 0.63  &1.82 & 7.17/2.21/0.77                 \\ 
         UG-Net (3-Channel)                   & 0.69  &1.86 & 7.89/2.31/0.81                  \\
         UG-Net (5-Channel)                   & 0.72  &1.94 & 8.51/2.43/0.83                  \\
         \bottomrule
    \end{tabular}
    \label{table:multichannel}
    \vspace{-0.5cm}
\end{table}

\begin{table}[]
\caption{Ablation study.}
\small
\centering
\setlength\tabcolsep{1.5pt} % default value: 6pt
    \begin{tabular}{p{5em} p{2em} p{6.5em} p{7em} p{2em}}
        \toprule
         Model                         & $D$                  &                                       &{SI-SNRi(dB)} /PESQ/STOI & RTF     \\
         \midrule               
\multirow{5}{5em}{UG-Net ($N=256$)}    &3                    & \multirow{3}{4.5em}{cLN \& No CI}   & 6.50/2.09/0.70         & 0.11    \\ 
                                       &4                    &                                      & 6.96/2.17/0.73        & 0.13    \\
                                       &5                    &                                      & 7.17/2.21/0.77        & 0.14    \\
                                \cmidrule{2-5}                   
                                       &\multirow{2}{2em}{5} &  cLN \& CI                          & 7.06/2.19/0.76         & 0.15    \\ 
                                \cmidrule{3-5}
                                       &                     &  LN \& No CI                        & 6.63/2.11/0.71         & 0.11    \\
 
         \bottomrule
         \end{tabular}
         \label{table:abalation}
\end{table}

\befSecSpace
\section{Conclusion}
\afSecSpace
This study presented a new architecture for varying-input-channel, time-domain, causal speech separation. Two novel mixer and separation modules were introduced to the TasNet system. The mixer module limits the model complexity while providing improved performance for increasing number of input channels. The separation module estimates masks of the different sources in a latent space employing a causal U-Net-like architecture inspired by the filter-and-process-based human auditory behavior. Experiments showed that using a single UX block in the separation module our system outperforms SOTA on time domain, causal speech separation while incurring lower computational and memory cost. Further increase in performance was observed when deepening the separation module with multiple successive UX blocks.\par

\befSecSpace
\section{Acknowledgement}
\afSecSpace
This work was supported by the National Institute on Deafness and Other Communication Disorders (NIDCD) of the National Institutes of Health (NIH) under Award 5R01DC015430-05. The content is solely the responsibility of the authors and does not necessarily represent the official views of the NIH.
% -------------------------------------------------------------------------
% \newpage
\bibliographystyle{IEEEbib}
% \bibliography{refer}
{\footnotesize
\bibliography{refer}}

\begin{thebibliography}{10}

\bibitem{speaker_ext}
Zbyn{\v{e}}k Koldovsk{\`y} and Petr Tichavsk{\`y},
\newblock ``Gradient algorithms for complex non-gaussian independent
  component/vector extraction, question of convergence,''
\newblock {\em IEEE Transactions on Signal Processing}, vol. 67, no. 4, pp.
  1050--1064, 2018.

\bibitem{bss}
Shoji Makino, Te-Won Lee, and Hiroshi Sawada,
\newblock {\em Blind speech separation}, vol. 615,
\newblock Springer, 2007.

\bibitem{supervised_overview}
DeLiang Wang and Jitong Chen,
\newblock ``Supervised speech separation based on deep learning: An overview,''
\newblock {\em IEEE/ACM Transactions on Audio, Speech, and Language
  Processing}, vol. 26, no. 10, pp. 1702--1726, 2018.

\bibitem{comprehensive_time_freq}
Fahimeh Bahmaninezhad, Jian Wu, Rongzhi Gu, Shi-Xiong Zhang, Yong Xu, Meng Yu,
  and Dong Yu,
\newblock ``A comprehensive study of speech separation: spectrogram vs waveform
  separation,''
\newblock {\em arXiv preprint arXiv:1905.07497}, 2019.

\bibitem{pit}
Dong Yu, Morten Kolb{\ae}k, Zheng-Hua Tan, and Jesper Jensen,
\newblock ``Permutation invariant training of deep models for
  speaker-independent multi-talker speech separation,''
\newblock in {\em 2017 IEEE International Conference on Acoustics, Speech and
  Signal Processing (ICASSP)}. IEEE, 2017, pp. 241--245.

\bibitem{tasnet}
Yi~Luo and Nima Mesgarani,
\newblock ``Tasnet: time-domain audio separation network for real-time,
  single-channel speech separation,''
\newblock in {\em 2018 IEEE International Conference on Acoustics, Speech and
  Signal Processing (ICASSP)}. IEEE, 2018, pp. 696--700.

\bibitem{convtasnet}
Yi~Luo and Nima Mesgarani,
\newblock ``Conv-tasnet: Surpassing ideal time--frequency magnitude masking for
  speech separation,''
\newblock {\em IEEE/ACM transactions on audio, speech, and language
  processing}, vol. 27, no. 8, pp. 1256--1266, 2019.

\bibitem{dprnn}
Yi~Luo, Zhuo Chen, and Takuya Yoshioka,
\newblock ``Dual-path rnn: efficient long sequence modeling for time-domain
  single-channel speech separation,''
\newblock in {\em ICASSP 2020-2020 IEEE International Conference on Acoustics,
  Speech and Signal Processing (ICASSP)}. IEEE, 2020, pp. 46--50.

\bibitem{DPTNN}
Jingjing Chen, Qirong Mao, and Dong Liu,
\newblock ``Dual-path transformer network: Direct context-aware modeling for
  end-to-end monaural speech separation,''
\newblock {\em arXiv preprint arXiv:2007.13975}, 2020.

\bibitem{GALR}
Max~WY Lam, Jun Wang, Dan Su, and Dong Yu,
\newblock ``Effective low-cost time-domain audio separation using globally
  attentive locally recurrent networks,''
\newblock in {\em 2021 IEEE Spoken Language Technology Workshop (SLT)}. IEEE,
  2021, pp. 801--808.

\bibitem{UNet}
Olaf Ronneberger, Philipp Fischer, and Thomas Brox,
\newblock ``U-net: Convolutional networks for biomedical image segmentation,''
\newblock in {\em International Conference on Medical image computing and
  computer-assisted intervention}. Springer, 2015, pp. 234--241.

\bibitem{waveUNet}
Daniel Stoller, Sebastian Ewert, and Simon Dixon,
\newblock ``Wave-u-net: A multi-scale neural network for end-to-end audio
  source separation,''
\newblock {\em arXiv preprint arXiv:1806.03185}, 2018.

\bibitem{SudoRmRf}
Efthymios Tzinis, Zhepei Wang, and Paris Smaragdis,
\newblock ``Sudo rm-rf: Efficient networks for universal audio source
  separation,''
\newblock in {\em 2020 IEEE 30th International Workshop on Machine Learning for
  Signal Processing (MLSP)}. IEEE, 2020, pp. 1--6.

\bibitem{DeepCASA}
Yuzhou Liu and DeLiang Wang,
\newblock ``Causal deep casa for monaural talker-independent speaker
  separation,''
\newblock {\em IEEE/ACM transactions on audio, speech, and language
  processing}, vol. 28, pp. 2109--2118, 2020.

\bibitem{brain-1}
Sanne Rutten, Roberta Santoro, Alexis Hervais-Adelman, Elia Formisano, and
  Narly Golestani,
\newblock ``Cortical encoding of speech enhances task-relevant acoustic
  information,''
\newblock {\em Nature human behaviour}, vol. 3, no. 9, pp. 974--987, 2019.

\bibitem{brain-2}
Liberty~S Hamilton, Yulia Oganian, Jeffery Hall, and Edward~F Chang,
\newblock ``Parallel and distributed encoding of speech across human auditory
  cortex,''
\newblock {\em Cell}, vol. 184, no. 18, pp. 4626--4639, 2021.

\bibitem{PReLU}
Kaiming He, Xiangyu Zhang, Shaoqing Ren, and Jian Sun,
\newblock ``Delving deep into rectifiers: Surpassing human-level performance on
  imagenet classification,''
\newblock in {\em Proceedings of the IEEE international conference on computer
  vision}, 2015, pp. 1026--1034.

\bibitem{DW_convolution}
Fran{\c{c}}ois Chollet,
\newblock ``Xception: Deep learning with depthwise separable convolutions,''
\newblock in {\em Proceedings of the IEEE conference on computer vision and
  pattern recognition}, 2017, pp. 1251--1258.

\bibitem{wsj}
John Garofolo, David Graff, Doug Paul, and David Pallett,
\newblock ``Csr-i (wsj0) complete ldc93s6a,''
\newblock {\em Web Download. Philadelphia: Linguistic Data Consortium}, vol.
  83, 1993.

\bibitem{librispeech}
Vassil Panayotov, Guoguo Chen, Daniel Povey, and Sanjeev Khudanpur,
\newblock ``Librispeech: An asr corpus based on public domain audio books,''
\newblock in {\em 2015 IEEE International Conference on Acoustics, Speech and
  Signal Processing (ICASSP)}, 2015, pp. 5206--5210.

\bibitem{deep-clustering}
John~R Hershey, Zhuo Chen, Jonathan Le~Roux, and Shinji Watanabe,
\newblock ``Deep clustering: Discriminative embeddings for segmentation and
  separation,''
\newblock in {\em 2016 IEEE international conference on acoustics, speech and
  signal processing (ICASSP)}. IEEE, 2016, pp. 31--35.

\bibitem{utterance-PIT}
Morten Kolb{\ae}k, Dong Yu, Zheng-Hua Tan, and Jesper Jensen,
\newblock ``Multitalker speech separation with utterance-level permutation
  invariant training of deep recurrent neural networks,''
\newblock {\em IEEE/ACM Transactions on Audio, Speech, and Language
  Processing}, vol. 25, no. 10, pp. 1901--1913, 2017.

\bibitem{habets-RIR}
Emanuel~AP Habets,
\newblock ``Room impulse response generator,''
\newblock {\em Technische Universiteit Eindhoven, Tech. Rep}, vol. 2, no. 2.4,
  pp. 1, 2006.

\bibitem{reverb}
Keisuke Kinoshita, Marc Delcroix, Takuya Yoshioka, Tomohiro Nakatani, Emanuel
  Habets, Reinhold Haeb-Umbach, Volker Leutnant, Armin Sehr, Walter Kellermann,
  Roland Maas, et~al.,
\newblock ``The reverb challenge: A common evaluation framework for
  dereverberation and recognition of reverberant speech,''
\newblock in {\em 2013 IEEE Workshop on Applications of Signal Processing to
  Audio and Acoustics}. IEEE, 2013, pp. 1--4.

\bibitem{backed-sdr}
Jonathan Le~Roux, Scott Wisdom, Hakan Erdogan, and John~R Hershey,
\newblock ``Sdr--half-baked or well done?,''
\newblock in {\em ICASSP 2019-2019 IEEE International Conference on Acoustics,
  Speech and Signal Processing (ICASSP)}. IEEE, 2019, pp. 626--630.

\bibitem{hungarian}
Harold~W Kuhn,
\newblock ``The hungarian method for the assignment problem,''
\newblock {\em Naval Research Logistics (NRL)}, vol. 52, no. 1, pp. 7--21,
  2005.

\bibitem{adam}
Diederik~P Kingma and Jimmy Ba,
\newblock ``Adam: A method for stochastic optimization,''
\newblock {\em arXiv preprint arXiv:1412.6980}, 2014.

\bibitem{PESQ}
Antony~W Rix, John~G Beerends, Michael~P Hollier, and Andries~P Hekstra,
\newblock ``Perceptual evaluation of speech quality (pesq)-a new method for
  speech quality assessment of telephone networks and codecs,''
\newblock in {\em 2001 IEEE international conference on acoustics, speech, and
  signal processing. Proceedings (Cat. No. 01CH37221)}. IEEE, 2001, vol.~2, pp.
  749--752.

\bibitem{STOI}
Cees~H Taal, Richard~C Hendriks, Richard Heusdens, and Jesper Jensen,
\newblock ``A short-time objective intelligibility measure for time-frequency
  weighted noisy speech,''
\newblock in {\em 2010 IEEE international conference on acoustics, speech and
  signal processing}. IEEE, 2010, pp. 4214--4217.

\end{thebibliography}

\end{document}